\documentclass[iop]{emulateapj}
\usepackage{color}
\usepackage{graphicx}
\usepackage{amsmath}

\shorttitle{constraining cosmological parameters}
\shortauthors{Liu et al.}

\begin{document}
\title{Implications from simulated strong gravitational lensing systems: constraining cosmological parameters using
Gaussian Processes }

\author{Tonghua Liu\altaffilmark{1}, Shuo Cao\altaffilmark{1$\dag$}, Jia Zhang\altaffilmark{2}, Shuaibo Geng\altaffilmark{1}, Yuting Liu\altaffilmark{1}, Xuan Ji\altaffilmark{1}, and Zong-Hong Zhu\altaffilmark{1}}

\altaffiltext{1}{Department of Astronomy, Beijing Normal University,
Beijing 100875, China; \emph{caoshuo@bnu.edu.cn}}
\altaffiltext{2}{Department of Physics, School of Mathematics and
physics, Weinan Normal University, Shanxi 714099, China}

\begin{abstract}

Strongly gravitational lensing systems (SGL) encodes cosmology
information in source/lens distance ratios $\mathcal{D}_{\rm
obs}=\mathcal{D}_{\rm ls}/\mathcal{D}_{\rm s}$, which can be used to
precisely constrain cosmological parameters. In this paper, based on
future measurements of 390 strong lensing systems from the
forthcoming LSST survey, we have successfully reconstructed the
distance ratio $\mathcal{D}_{\rm obs}$ (with the source redshift
$z_s\sim 4.0$), directly from the data without assuming any
parametric form. A recently developed method based on
model-independent reconstruction approach, Gaussian Processes (GP)
is used in our study of these strong lensing systems. Our results
show that independent measurement of the matter density parameter
($\Omega_m$) could be expected from such strong lensing statistics.
More specifically, one can expect $\Omega_m$ to be estimated at the
precision of $\Delta\Omega_m\sim0.015$ in the concordance
$\Lambda$CDM model, which provides comparable constraint on
$\Omega_m$ with Planck 2015 results. In the framework of the
modified gravity theory (DGP), 390 detectable galactic lenses from
future LSST survey would lead to stringent fits of
$\Delta\Omega_m\sim0.030$. \textbf{Finally, we have discussed three
possible sources of systematic errors (sample incompleteness, the
determination of length of lens redshift bin, and the choice of lens
redshift shells), and quantified their effects on the final
cosmological constraints. Our results strongly indicate that future
strong lensing surveys, with the accumulation of a larger and more
accurate sample of detectable galactic lenses, will considerably
benefit from the methodology described in this analysis. }

\end{abstract}


\maketitle


\section{Introduction}

During the last decades, one of the most important issues of modern
cosmology is the accelerating expansion of the universe, which has
been discovered and verified by several observational probes
including the type Ia supernova (SNe Ia)
\citep{Riess,Perlmutter,Riess04,Knop}, baryon acoustic oscillation
(BAO) \citep{Percival}, and precise measurements of the spectrum of
cosmic microwave background (CMB)
\citep{Balbi,Jaffe,Spergel03,Spergel07}. Currently, the detailed
dynamics of the accelerated expansion is still not well known. The
origin of this acceleration may be attributed to dark energy with
negative pressure, based on the cosmological principles
(homogeneous, isotropic) and Einstein¡¯s general relativity (GR). In
the framework of the current standard model, the so-called
$\Lambda$CDM model, the accelerated cosmological expansion is
powered by Einstein's cosmological constant, $\Lambda$, a spatially
homogeneous fluid with equation of state parameter $w=p/\rho=-1$
(with $p$ and $\rho$ being the fluid pressure and energy density).
However, one should note that the $\Lambda$CDM model, although
providing a reasonable fit to most observational constraints, is
still confronted with the well-known coincidence problem and
fine-tuning problem \citep{Weinberg}. See \citet{Cao11a,Cao14} and
references therein for recent discussions about more dark energy
models under discussion \citep{Cao11b,Cao13,Cao15a,Qi18} .

On the other hand, dark energy is not the only possible explanation
of the present cosmic acceleration, and it is argued that the
observed accelerated expansion should instead be viewed as the
possible deviation from Einstein's theory of gravity on large
cosmological length scales. For instance, some unknown physical
processes involving modifications of gravity theory can also account
for this apparently unusual phenomenon. Some modifications are
related to the possible existence of extra dimensions, which gives
rise to the so-called braneworld cosmology. In this paper we
investigate constraints on one interesting braneworld cosmological
model proposed by \citet{Dvali00a,Arkani,Dvali00b}, the Dvali-Gabadadze-Porrati
(DGP) braneworld, which is often used to describe a gravity spilling
over large scales and into higher dimensions. So far, both models
derived from introducing an exotic component like dark energy and
those established by modifying Einstein's theory of gravity can
survive the above-mentioned observations. Actually, the
investigation of the expected constraints on DGP braneworld
cosmology has been performed from different astrophysical
observations \citep{Xu10,Giannantonio,Lombriser,Wang08}. However, it is interesting to
note that based on different theoretical basis, the determination of
the same cosmological parameter in different cosmological models are
clearly different. The normal branch of DGP gravity is confronted by
the currently available cosmic observations from the geometrical and
dynamical perspectives. For instance, ref.~\citet{Xu14} made a joint
analysis of the DGP braneworld cosmology with the Supernova Legacy
Survey (SNLS) data, first released CMB data from Planck, and
redshift space distortion (RSD) data ($\Omega_m=0.286\pm 0.008$).
While comparing the results with those obtained from Planck 2018
data (TT, TE, EE+lowE+lensing) based $\Lambda$CDM model
$\Omega_m=0.315\pm0.007$ \citep{Aghanim}, differences in central values
of the best-fit cosmological parameter were clearly reported.
Similar analyses were carried out by \citet{Ma19}. If one wants to
place more comprehensive cosmological constraints on a possible
model or distinguish between dark energy and modified gravity
theories, it is crucial to measure the expansion rate of universe at
many different redshifts.

The power of modern cosmology lies in building up consistency rather
than in single, precise, crucial experiments, which implies that
every alternative method of restricting cosmological parameters is
desired. In particular, a new cosmological window would open if we
could measure the cosmic expansion directly within the "redshift
desert", roughly corresponding to redshifts $2<z<5$. As one of the
successful predictions of general relativity in the past decades,
strong gravitational lensing has become a very important
astrophysical tool allowing us to use individual lensing galaxies to
measure cosmological parameters \citep{Treu06}. When the source, lens,
and observer are sufficiently well aligned, the deflection of light
forms an Einstein ring, from which the source/lens distance ratios
can be obtained. \citet{Biesiada06} first proposed the possible
application of this kind of observation as a cosmological tool, the
importance of which method was stressed again by \citet{Grillo,Biesiada}. The
idea of using such systems for measuring the cosmic equation of
state was discussed in \citet{Cao12JC} and also in a more recent paper
by \citet{Cao15Ap}. The angular diameter distance ratios may also be
used to constrain different cosmological parameters in various
cosmological models \citep{Futamase,Treu640,Melia}. On the one hand, in order to
achieve high precision constraints on the cosmological parameters,
it is still necessary to develop new complementary techniques
bridging the redshift gap of current data, and furthermore increase
the depth and quality of observational data sets. In this paper, we
will use the model-independent method Gaussian processes (GP) to
reconstruct one-dimensional function of the angular diameter
distance ratios, with fixed lens (or source) redshift. An obvious
benefit of this approach is that GP allow one to reconstruct a
function from data directly without any parametric assumption, which
has been widely used in various studies \citep{Seikel12a,Seikel12b,Cai,Yennapureddy,Melia18b}.
The first (to our knowledge) formulations of this approach can be
traced back to \citet{Yennapureddy}, which revisited the most recent and
significantly improved observations of early-type gravitational
lenses (158 combined systems) to distinguish $\Lambda$CDM another
Friedmann-Robertson-Walker (FRW) cosmology known as the $R_h=ct$
universe. Their results showed that, the probability of $R_h=ct$
(which is characterized by a total equation of state $w=-1/3$) being
the correct cosmology is higher than that of $\Lambda$CDM, with a
degree of significance that grows with the number of sources
considered. Therefore, although the differentiation of competing
cosmologies is already quite competitive compared with those from
other methods, it still suffer from the small number of lenses in
the statistical sample.

In the near future, the next generation of wide and deep sky
surveys, with improved depth, area and resolution may increase the
current galactic-scale lens sample sizes by orders of magnitude. The
purpose of our paper is to investigate the constraining capability
of SGL on some fundamental cosmological parameters, using the
simulated SGL sample based on the forthcoming Large Synoptic Survey
Telescope (LSST) survey. More importantly, compared with the
previous procedure of carrying out the reconstruction within thin
redshift-shells of sources \citep{Yennapureddy}, we turn $D_{ls}/D_s$ into a
one-dimensional function of source redshift ($z_s$) for what is
essentially a fixed lens redshift ($z_l$). The advantage of this
work lies in the fact that, we could achieve reasonable constraints
on cosmological parameters at much higher redshifts ($z\sim4$), when
the sample is large enough to yield enough statistics to warrant
this approach. As can clearly seen from the previous analysis
\citep{Yennapureddy}, the current SGL sample is not sufficient enough to
extend our investigation to $z\sim 4$ (the data are less dispersed
in the lens plane, and scattered much more in the source plane).

This paper is organized as follows. In Sect.~2 we briefly describe
the methodology and the simulated strong lensing data from LSST. In
Sect.~3 we introduce our improved Gaussian processes and the area
minimization statistic. Two prevalent cosmologies and the fitting
results on the relevant cosmological parameters are presented in
Sect.~4. Finally, we summarize our conclusions in Sect.~5.

\section{Simulated strong lensing systems}

For a specific strong lensing system with the intervening galaxy
acting as a lens (at redshift $z_l$), the multiple image separation
of the source (at redshift $z_s$) depends only on angular diameter
distances to the lens and to the source, as long as one has a
reliable model for the mass distribution within the lens. Moreover,
compared with late-type and unknown-type counterparts, early-type
galaxies are more likely to serve as intervening lenses for the
background sources (quasars or galaxies). This is because such
galaxies contain most of the cosmic stellar mass of the Universe.
The recently released large sample include 118 galaxy-scale strong
gravitational lensing systems discovered and observed in SLACS,
BELLS, LSD, and SL2S surveys \citep{Cao15Ap}, which can be used to
place stringent constrains on cosmological parameters in alternative
cosmological models \citep{Li16}, and to study the mass density
distribution in early-type galaxies \citep{Cao16}. Recent analytical
work has forecast the number of galactic-scale lenses to be
discovered in the forthcoming photometric surveys \citep{Collett15}.
Such a significant increase of the number of strong lensing systems
will considerably improve the constraints on the cosmological
parameters. With a large increase to the known strong lens
population, current work could be extended to a new regime: what
kind of cosmological results one could obtain from $\sim$10,000
discoverable lens population in the forthcoming Large Synoptic
Survey Telescope (LSST) survey?

Using the simulation programs publicly available
\footnote{github.com/tcollett/LensPop}, we carry out a Monte Carlo
simulation of the lens and source populations to forecast the yields
of LSST. In our simulation, 10000 SGL systems has been obtain with
the proper inputs in the following three assumptions: (i) early-type
galaxies act as lenses; (ii) mass distribution of lens is
approximated by the power law model; (iii) the normalization and
shape of the velocity dispersion function of early-galaxies are not
varying with redshift. The assumptions are well consistent with the
previous studies on lensing statistics if all galaxies are
early-type \citep{Chae,Mitchell,Capelo}. Moreover, we assume a flat
concordance $\Lambda$CDM model with $\Omega_m=0.30$ as a fiducial
cosmology.

\begin{figure}
\begin{center}
\includegraphics[width=0.9\linewidth]{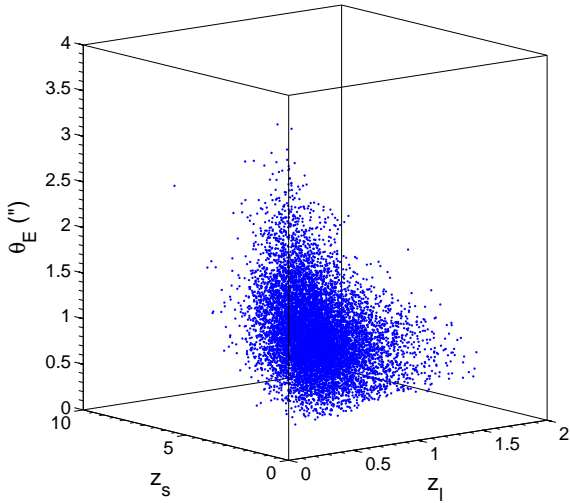}
\end{center}
\caption{Scatter plot of 10000 simulated strong lensing systems from
future LSST survey.}
\end{figure}

Motivated by several previous studies supporting that early-type
galaxies are well described by power-law mass distributions in
regions covered by the X-ray and lensing observations
\citep{Humphrey10,Koopmans06}, we model the lens galaxy with a
power-law mass distribution ($\rho \sim r^{- \gamma}$). The main
idea of our method is that formula for the Einstein radius in a
power-law lens expresses as
\begin{equation} \label{Einstein} \theta_E =   4 \pi
\frac{\sigma_{ap}^2}{c^2} \frac{D_{ls}}{D_s} \left(
\frac{\theta_E}{\theta_{ap}} \right)^{2-\gamma} f(\gamma),
\end{equation}
based on which the ratio of angular-diameter distances between lens
and source ($D_{ls}$) and between observer and source ($D_{s}$) can
be obtained
\begin{equation} \label{Einstein}  \frac{D_{ls}}{D_s} =   \frac{\theta_E} {4 \pi}
\frac{c^2}{\sigma_{ap}^2}  \left( \frac{\theta_E}{\theta_{ap}}
\right)^{\gamma-2} f(\gamma)^{-1},
\end{equation}
$f(\gamma)$ represents a function of the radial mass profile slop
\citep{Koopmans} and $\sigma_{ap}$ is the luminosity averaged
line-of-sight velocity dispersion of the lens inside the aperture
radius, $\theta_{ap}$ (more precisely, luminosity averaged
line-of-sight velocity dispersion). It is obvious that combining
$\sigma_{ap}$, $\theta_E$, $\theta_{ap}$ and $\gamma$ obtained from
the observations will introduce the measurement of the distance
ratio of $D_{ls}/D_s$. Current observational techniques allow the
redshifts of the lens $z_l$ and the source $z_s$ to be measured
precisely. Moreover, imaging and spectroscopy from the Hubble Space
Telescope (HST) and ground-based observatories make it possible to
derive two key ingredients for individual lenses: stellar velocity
dispersion, high-resolution images of the lensing systems. We take
the fractional uncertainty of the Einstein radius at a level of 1\%,
which is reasonable for the future LSST survey to obtain
high-resolution imaging with different stacking strategies for
combining multiple exposures \citep{Collett16}. Following the Lens
Structure and Dynamics (LSD) survey and the more recent Sloan Lenses
ACS (SLACS) survey, we take the fractional uncertainty of the
observed velocity dispersion at a level of 5\%, which can be
assessed from the spectroscopic data for central parts of lens
galaxies. More importantly, it was shown that the power-law mass
profile is still a useful assumption in gravitational lensing
studies and should be accurate enough as first-order approximations
to the mean properties of galaxies relevant to statistical lensing
(lenses observed in different surveys with the following median
values of the lens redshifts: $z_l = 0.215$ for SLACS, $z_l = 0.517$
for BELLS, $z_l = 0.81$ for LSD and $z_l = 0.456$ for SL2S
\citep{Cao15Ap}). In our fiducial model, the average logaritmic density
slope is modeled as $\gamma=2.085$ with 10\% intrinsic scatter, the
results from SLACS strong-lens early-type galaxies with direct
total-mass and stellar-velocity dispersion measurements
\citep{Koopmans09}.

\begin{figure}
\begin{center}
\includegraphics[width=0.9\linewidth]{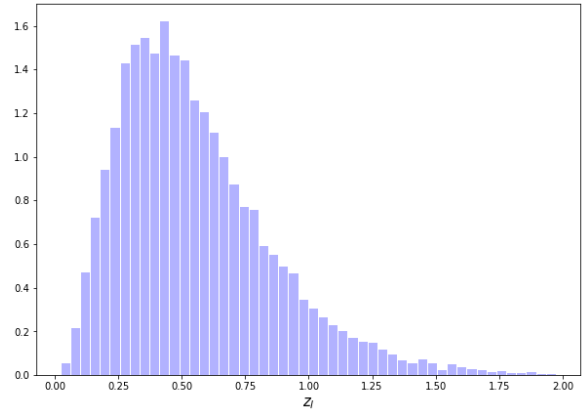}
\end{center}
\caption{\textbf{The lens redshift distribution of simulated strong
lensing systems from future LSST survey.}}
\end{figure}

Following the LSST observation simulator with the assumed survey
parameters summarized in Table 1 of \citet{Collett15}, we have
generated a realistic population of galaxies lensed by early-type
galaxies, assuming distributions of velocity dispersions and
Einstein radii similar to the SL2S sample \citep{Sonnenfeld13}. The
velocity dispersion function of the lenses in the local Universe
follows the modified Schechter function \citep{Sheth03}
\begin{equation} \label{stat2} \frac{d n}{d \sigma}= n_*\left(
\frac{\sigma}{\sigma_*}\right)^\alpha \exp \left[ -\left(
\frac{\sigma}{\sigma_*}\right)^\beta\right] \frac{\beta}{\Gamma
(\alpha/\beta)} \frac{1}{\sigma} \, , \end{equation} where $\alpha$
is the low-velocity power-law index, $\beta$ is the high-velocity
exponential cut-off index, $n_*$ is the integrated number density of
galaxies, and $\sigma_*$ is the characteristic velocity dispersion.
In this paper, we use the measurement of velocity dispersion
function (VDF) for local early-type galaxies, based on the much
larger SDSS Data Release 5 data set \citep{Choi07}. See
\citet{Cao12b} for discussion about such choice in view of other
data on velocity dispersion distribution functions. Currently, it
was found that simple evolutions do not significantly affect the the
appealing results based on lensing statistics, especially those from
the early-type galaxy number counts \citep{Im02} and the redshift
distribution of early-type lens galaxies \citep{Ofek03}. Therefore,
it is assumed in our analysis that the normalization and shape of
the velocity dispersion function of early-galaxies are not varying
with redshift. The population of strong lenses is dominated by
galaxies with velocity dispersion of $\sigma_{ap}=210\pm50$ km/s,
while the lens redshift distribution is well approximated by a
Gaussian with mean 0.40. Although discovering strong lenses in these
surveys will require the development of new methods and algorithms
\citep{Gavazzi14}, which is beyond the scope of this paper, we are
confident that the simulated population of lenses is a good
representation of what the future LSST survey might yield
\citep{Sonnenfeld13,Gavazzi14}.

The scatter plot of the simulated lensing systems is shown in
Fig.~1, from which one can see the LSST lenses result in a fair
coverage of lenses and sources redshifts.

\begin{figure}
\begin{center}
\includegraphics[width=0.9\linewidth]{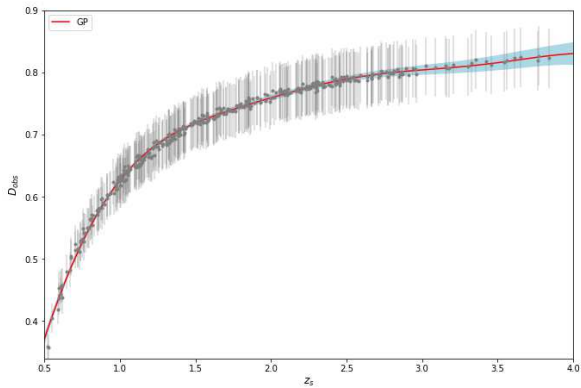}
\end{center}
\caption{The solid red curve in plot indicates the reconstructed
$\mathcal{D}_{\rm obs}$ function using Gaussian processes, for the
lens redshift ranges $0.30\sim0.32$. The light blue represents the
$1\sigma$ confidence region.}
\end{figure}

\section{Gaussian Processes}

In order to reconstruct the evolution of angular diameter distances
from simulated SGL data sets, we should find a model-independent
method to reconstruct $D=D_{ls}/D_s$. Although there are several
methods such as principle component analysis \citep{Huterer03} and
Gaussian smoothing \citep{Shafieloo06}, in this paper we will
reconstruct $D_{ls}/D_s$ more precisely by using the Gaussian
processes (GP) method.

A model-independent method of Gaussian processes \citep{Seikel12a},
can be employed to reconstruct the angular-diameter distance ratio
from the strong lensing data straightforwardly, without any
parametric assumption regarding cosmological model. Such approach
has been used in various studies in the literature
\citep{Qi19a,Cao19}. The distribution over functions provided by GP
is suitable to describe the observed data. Not that for each
lens-source pairing, $D_{ls}/D_s$, two angular diameter distances
are involved. Therefore, a reconstruction of $D_{ls}/D_s$ in two
dimensions is required, which is very difficult to handle with GP.
Following the methodology proposed by \citet{Yennapureddy}, one
interesting solution of this problem is to consider small redshift
ranges (with fixed source/lens redshift), which may effectively
reduce the problem to a one-dimensional reconstruction. In this
work, we choose to carry out the reconstruction within thin
redshift-shells of lenses, which makes it possible to turn
$D_{ls}/D_s$ into a one-dimensional function of source redshift
($z_s$) and thus achieve reasonable constraints on cosmological
parameters at much higher redshifts ($z\sim4$). Note that for the
purpose of GP reconstruction in one dimension, we assume that all
the lenses in redshift bin $(z_l,z_l+\Delta z)$ have the same
average redshift $z_l+\Delta z_l/2$. In order to minimize the
scatter in lens redshifts, we use a bin size less than 0.02 ($\Delta
z_l=0.02$). We also find that the choice of $\Delta z_l$ may play an
important role in the accuracy and reliability of our test, which
will be discussed in Section 4. More importantly, in order to
guarantee the precision of GP reconstruction, the selected
sub-sample should be large enough to yield enough statistics. In our
simulated sample of strong-lensing systems, these criteria therefore
allow us to assemble a sub-sample including 390 strong-lensing
systems, with the lensing galaxies covering the redshift shell of
$0.3<z_l<0.32$. \textbf{Fig.~2 shows the lens redshift distribution
of galactic-scale lenses discoverable in forthcoming LSST surveys.
It is apparent that, compared to the current surveys with the
following median values of the lens redshifts: SLACS -- $z_l =
0.215$, BELLS -- $z_l = 0.517$, LSD -- $z_l = 0.81$ and SL2S -- $
z_l = 0.456$ \citep{Cao15Ap}, the future LSST survey is particularly
promising to discover more lenses covering the redshift range of
$0.25 - 0.50$. Therefore, the thin shell of $0.3 < z_l < 0.32$ is a
good statistical representation of the simulated population of
lenses what the future LSST survey might yield.}

At each point $z$, the reconstructed function $f(z)$ is also a
Gaussian distribution with a mean value and Gaussian error. In this
process, the values of the reconstructed function evaluated at any
two different points $z$ and $\tilde{z}$, are connected by a
covariance function $k(z,\tilde{z})$. In this paper, we take the
Mat\'{e}rn ($\nu = 9/2$) covariance function
\begin{align}
k(z,\tilde z) = &~{\sigma _f}^2\exp\left( - \frac{{3\left| {z - \tilde z} \right|}}{\ell }\right) \nonumber \\
      &~~\times\Big[1 + \frac{{3\left| {z - \tilde z} \right|}}{\ell } + \frac{{27{{(z - \tilde z)}^2}}}{{7{\ell ^2}}} \nonumber \\
     &~~ + \frac{{18{{\left| {z - \tilde z} \right|}^3}}}{{7{\ell ^3}}} + \frac{{27{{(z - \tilde z)}^4}}}{{35{\ell ^4}}}\Big],
\end{align}
where $\ell$ provides a measure of the coherence length of the
correlation in $x$-direction and $\sigma_f$ is the overall amplitude
of the correlation in the $y$-direction. The values of the two hyper
parameters $\sigma_f$ and $\ell$ will be optimized by GP with the
observed data set. This implies that the reconstructed function is
not dependent on the initial hyper-parameter settings, which
guarantees the reliability of the reconstructed function. Compared
with the squared exponential form for covariance function, which has
been widely used in the literature
\citep{Seikel12a,Seikel12b,Cai,Yennapureddy,Melia18b}, the
Mat\'{e}rn ($\nu=9/2$) covariance function can lead to more reliable
results when applying GP to reconstructions using distance
measurements \citep{Yang15}. Using this covariance function, values
of data points at other redshifts which have not be observed can
also be obtained, which could effectively bridges the redshift gap
between current data. Following \citet{Seikel12a} in which the
detailed technical description of GP can be found, we use the
Gaussian processes in Python (GaPP) \footnote{
http://www.acgc.uct.ac.za/seikel/GAPP/index.html} to execute the
model-independent method and derive our GP results. The
reconstructed function $\mathcal{D}_{\rm obs}(\langle z_l\rangle,
z_s)$, as well as the estimation of the $1\sigma$ confidence region
with the 390 simulated strong lensing systems is shown in Fig.~3.

In order to demonstrate how the reconstructed function
$\mathcal{D}_{\rm obs}(\langle z_l\rangle, z_s)$ works, we have
constrained two simple cosmological models: the $\Lambda$CDM and DGP
models under assumption of spatially flat Universe. On the other
hand, in the face of different competing cosmological scenarios, it
is important to find an effective way to decide which one is most
favored by the data. Following the analysis of \citet{Yennapureddy}, a new
type of model comparison statistics, Area Minimization Statistics,
will be used for this purpose.

\section{Competing cosmological models and statistical analysis}

Flatness of the Friedmann-Robertso-Walker (FRW) metric is assumed in
our analysis, which is strongly supported by the recent Planck
results \citep{Ma19} and independently supported by the observations
of milliarcsecond compact structure of radio quasars at $z\sim 3.0$
\citep{Cao17,Cao19}. In a zero-curvature universe filled with
ordinary pressureless matter (cold dark matter plus baryons), dark
energy, and negligible radiation, the Friedmann equation reads
\begin{equation}
H^2(z)=H_0^2[\Omega_m(1+z)^3+(1-\Omega_{m})],
\end{equation}
where $\Omega_{m}$ is the current density fraction of matter
component. In the framework of this standard cosmological model, the
angular diameter distance between redshifts $z_1$ and $z_2$ becomes
\begin{eqnarray}
D_A^{\Lambda CDM}(z_1,z_2)&=&\frac{c}{H_0 (1+z_2)}\times\nonumber\\
\null&\null&\hskip-0.3in
\int^{z_2}_{z_1}\frac{dz}{\sqrt{\Omega_m(1+z)^3+1-\Omega_m}}.\nonumber
\end{eqnarray}
Over the past decades, the importance of modified gravity theories
was stressed again. The DGP model \citep{Dvali00b}, which accounts for the
cosmic acceleration without dark energy, arises from a class of
brane world theories in which gravity leaks out into the bulk at
large distances. More specifically, this leaking of gravity takes
place only above a certain cosmological scale $r_c$. In the
framework of a spatially flat DGP model, the Friedmann equation can
be expressed as
\begin{equation}
H^2-\frac{H}{r_c}=\frac{8\pi G}{3}\rho_m,
\end{equation}
where the length at which the leaking occurs can be associated with
the density parameter: $\Omega_{r_c}=1/(4r_c^2H_0^2)$. It is also
straightforward to check the validity of the relation
$\Omega_{r_c}=\frac{1}{4}(1-\Omega_m)^2$ in the flat DGP model.
Thus, in the framework of a spatially flat DGP model, we can
directly rewrite the above equation and obtain the angular diameter
distance between redshifts $z_1$ and $z_2$
\begin{eqnarray}
D_A^{DGP}(z_1,z_2)&=&\frac{c}{H_0 (1+z_2)}\times\nonumber\\
\null&\null&\hskip-0.3in
\int^{z_2}_{z_1}[\sqrt{\Omega_m(1+z)^3+\Omega_{r_c}}+\sqrt{\Omega_{r_c}}]^{-1}
dz.\nonumber
\end{eqnarray}

Now we introduce a new statistic, the ``Area Minimization Statistic"
to constrain cosmological parameters, which has been recently
proposed to test the evolution of the Universe, and then applied to
the investigation of dynamical properties of dark energy
\citep{Yennapureddy17,Melia18c}. It should be noted that our reconstructed distance
ratio $\mathcal{D}_{\rm obs}$, with the corresponding theoretical
model value $\mathcal{D}^{\rm \Lambda{\rm CDM}}$ or
$\mathcal{D}^{\rm DGP}$, is an continuous function. Therefore, the
discrete sampling statistics (such as the $\chi^2$ statistic) is not
sufficient enough to provide an effective way to make a comparison
between different models, because the sampling at random points to
obtain the squares of differences between model and reconstructed
curve would lose information between these points.

\begin{figure}
\begin{center}
\includegraphics[width=0.9\linewidth]{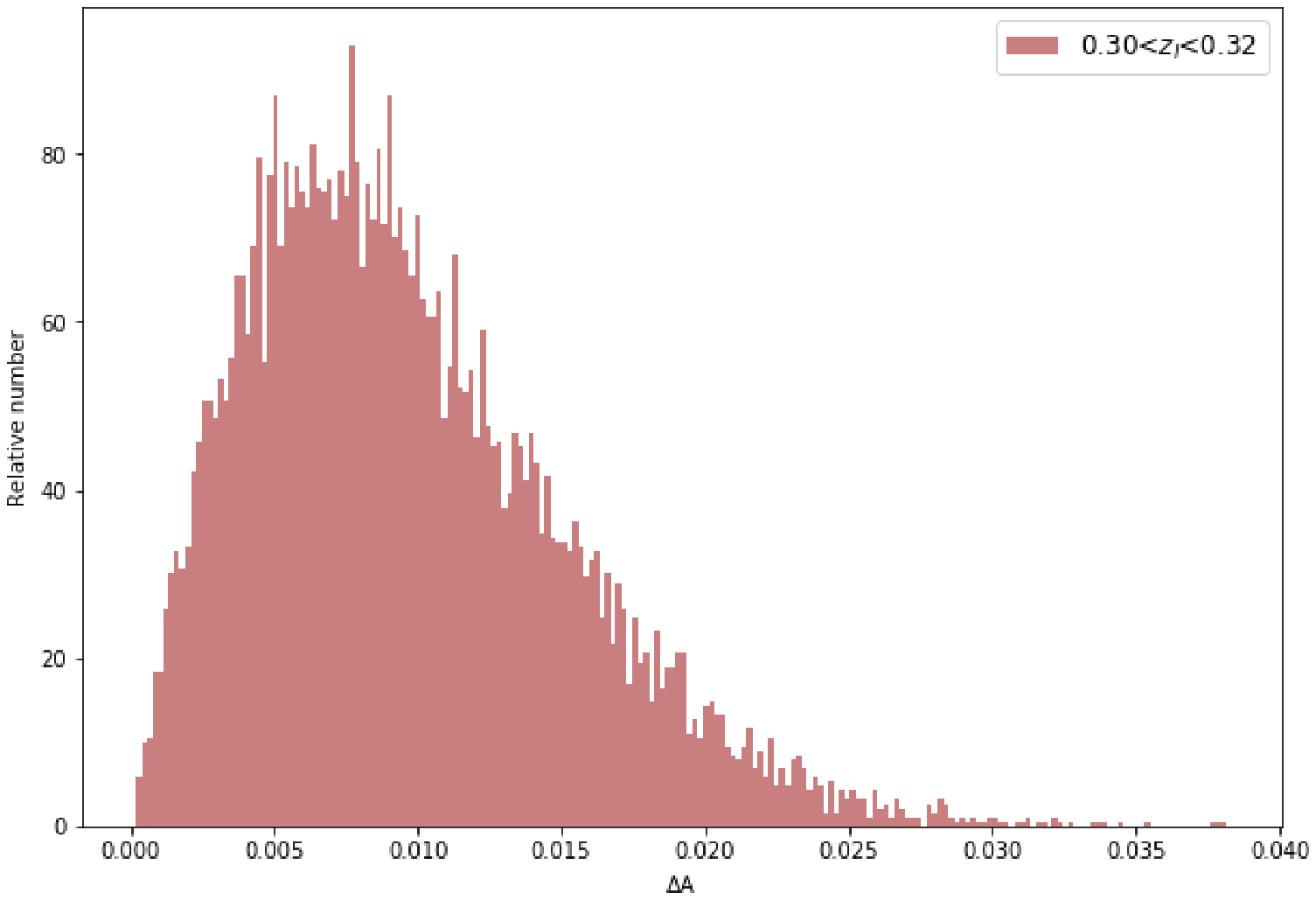}\\
\vskip 0.2in
\includegraphics[width=0.9\linewidth]{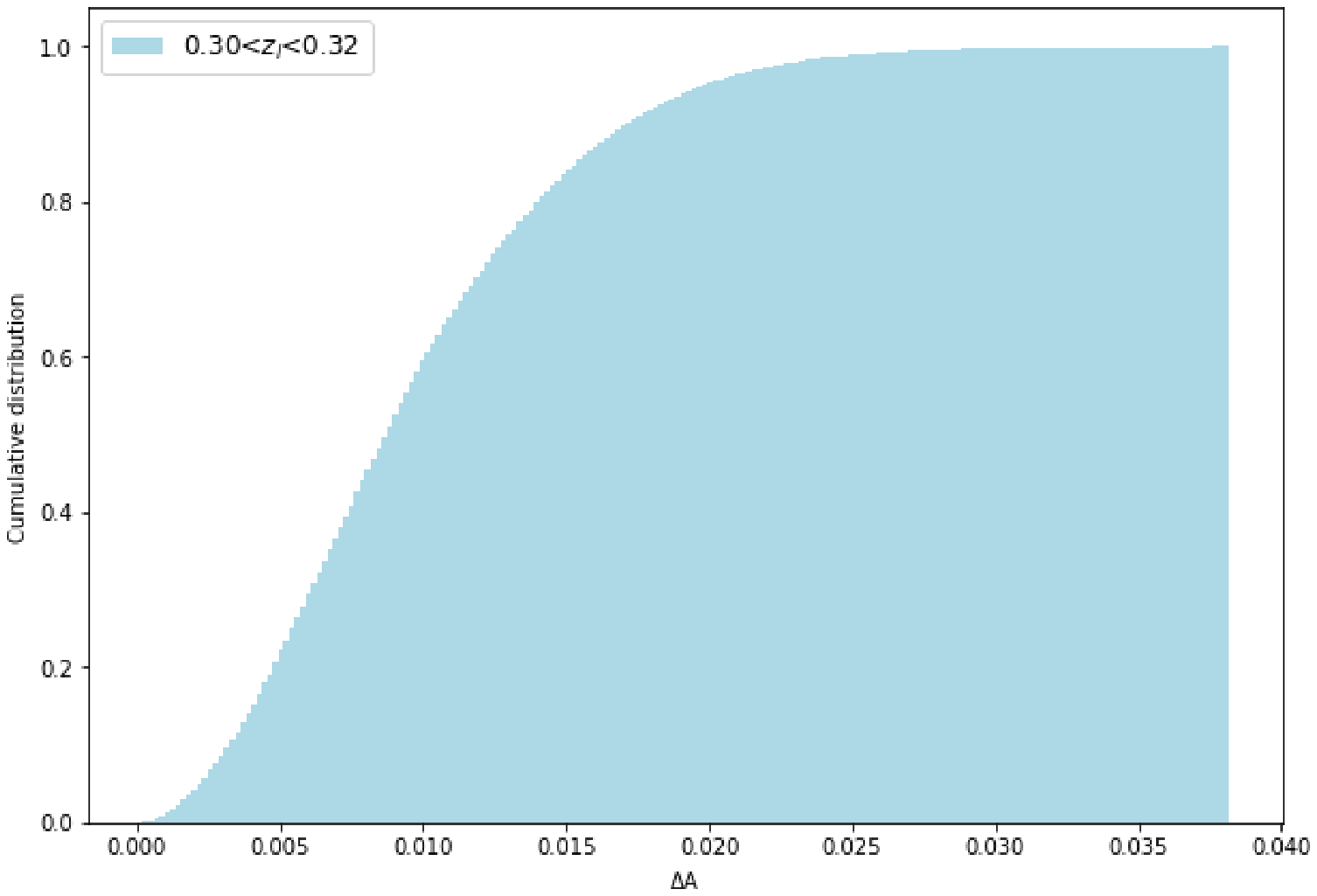}
\end{center}
\caption{{\it Top panel:} The distribution of frequency versus area
differential $\Delta A$ for a mock sample with lens bin
$0.3<z_l<0.32$;  {\it Bottom panel:} its corresponding cumulative
probability distribution.}
\end{figure}

The most important assumption of ``Area Minimization Statistic" is
that the measurement errors are Gaussian, which should be satisfied
by the mock sample with GP reconstructed curves and possible
variation of $\mathcal{D}$ away from $\mathcal{D}_{\rm obs}$. More
specifically, such statistic is realized by a Gaussian randomized
value
\begin{equation}
\mathcal{D}_{i,\,{\rm mock}}(\langle z_l\rangle, z_s)=\mathcal{D}_{i,\,{\rm obs}}(\langle z_l\rangle, z_s)+r
\sigma_{\mathcal{D}_{i,obs}}\;,
\end{equation}
where $r$ is characterized by a Gaussian distribution $r=0.0\pm1.0$,
and $\mathcal{D}_{i,\,{\rm obs}}(\langle z_l\rangle, z_s)$
represents the actual measurement at source redshift $z_s$, with
1$\sigma$ error denoted by $\sigma_{\mathcal{D}_{i,obs}}$.
Therefore, the function $\mathcal{D}_{\rm mock}(\langle z_l\rangle,
z_s)$ corresponding to mock sample could be straightforwardly
obtained. Finally, a normalized absolute area difference between
$\mathcal{D}_{\rm mock}(\langle z_l\rangle, z_s)$ and the GP
reconstructed function of the actual data can be defined as
\begin{equation}
\Delta A= \int_{z_{\rm min}}^{z_{\rm
max}}dz_s\bigg(\frac{\big|\mathcal{D}_{\rm mock}(\langle z_l\rangle,
z_s)-
        \mathcal{D}_{\rm obs}(\langle z_l\rangle, z_s)\big|}{\sigma_{\mathcal{D}}}\bigg)\;,
\end{equation}
where $z_{\rm min}$ and $z_{\rm max}$ are the minimum and maximum
redshifts of the mock sample. This process is repeated 10000 times
in order to guarantee unbiased final results, from which one could
derive a distribution of frequency versus area differential $\Delta
A$ and the cumulative probability distribution. The results are
shown in Fig.~4, in which one can clearly see a 1-to-1 mapping
between the value of $\Delta A$ and the corresponding frequency.
More importantly, the cumulative distribution for a given $\Delta A$
quantifies the fraction of the randomized realizations whose
differential area is smaller than this value.

In the framework of a specific cosmological model, we can calculate
a normalized absolute area difference between the GP reconstructed
function of the actual data and its theoretical counterpart
\begin{equation}
\Delta A= \int_{z_{\rm min}}^{z_{\rm
max}}dz_s\bigg(\frac{\big|\mathcal{D}_{\rm mock}(\langle z_l\rangle,
z_s)-\mathcal{D}_{\rm th}(\langle z_l\rangle,
z_s)\big|}{\sigma_{\mathcal{D}}}\bigg)\;.
\end{equation}
Therefore, based on the assumption that a curve with a smaller
$\Delta A$ is a better match to $\mathcal{D}_{\rm obs}$, the
cumulative distribution can be directly used to estimate the
probability (i.e., the p-value) that a cosmological model is well
consistent with the observations. More specifically, in order to
decide which cosmology is favored by the observational data, we
perform model comparison statistics by calculating its $\Delta A$
and apply the 1-to-1 mapping to determine the probability that it is
inconsistent with the SGL sample. The results for different
cosmological scenarios on the reconstructed $\mathcal{D}$
observations are listed in Table 1 and discussed as follows. We
stress here that the observational distance ratio $\mathcal{D}$ has
the advantage that the Hubble constant $H_0$ gets cancelled, hence
it does not introduce any uncertainty to the results.

\begin{figure}
\begin{center}
\includegraphics[width=0.9\linewidth]{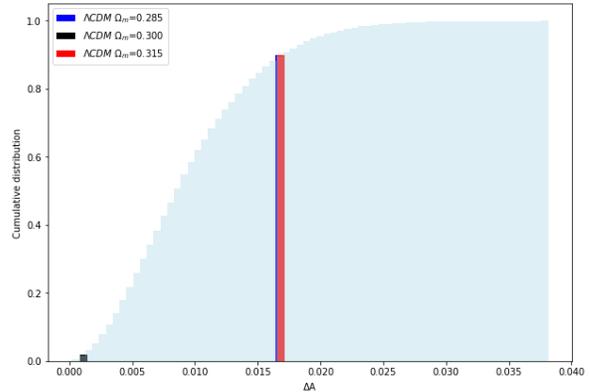}
\end{center}
\caption{ The cumulative probability distributions with the matter
parameters $\Omega_m=0.285$ (blue), $\Omega_m=0.300$ (black) and
$\Omega_m=0.315$ (red) in $\Lambda$CDM cosmology.}
\end{figure}

\begin{figure}
\begin{center}
\includegraphics[width=0.9\linewidth]{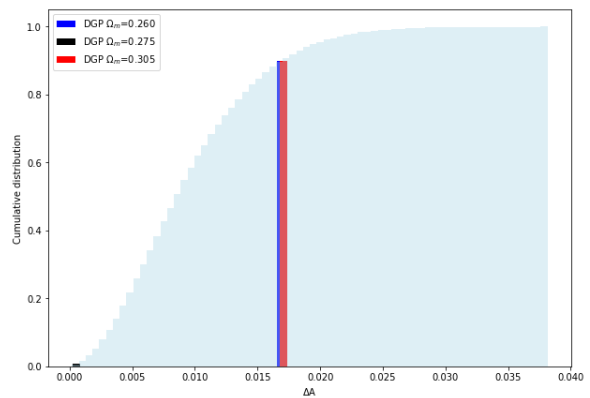}
\end{center}
\caption{ The corresponding cumulative probability distributions
with the matter parameters $\Omega_m=0.260$ (blue), $\Omega_m=0.275$
(yellow) and $\Omega_m=0.305 $(red) in DGP cosmology.}
\end{figure}

\begin{table}
\begin{center}
\caption{Summary of the cosmological constraints using strong
gravitational lenses with Gaussian Processes.}
\begin{tabular}{c|l|l}\hline\hline
Cosmological model     & Cosmological parameter & Probability  \\
\hline
$\Lambda$CDM      & $\Omega_m=0.285$ & $10.00\%$   \\
~~~~~~~~~~~   & $\Omega_m=0.300$&    $99.99\%$ \\
~~~~~~~~~~~     & $\Omega_m=0.315$ &   $10.00\%$  \\

  \hline

 DGP   &  $\Omega_m=0.260$ &$10.00\%$    \\

 ~~~~~~~~~~~   &  $\Omega_m=0.275 $  &$99.99\%$    \\
~~~~~~~~~~~   &    $\Omega_m=0.305 $   &$10.00\%$  \\

\hline\hline
\end{tabular}
\end{center}
\end{table}

We start our analysis with the $\Lambda$CDM model with constant dark
energy density and constant cosmic equation of state $w=-1$. The
corresponding cumulative probability distributions are plotted in
Fig.~5, which also locate the $\Delta A$ values with different
matter density parameter: $\Omega_m=0.300$ (black), $\Omega_m=0.285$
(blue), and $\Omega_m=0.315$ (red). The probabilities associated
with these differential areas are summarized in Table~1. As can be
clearly seen from the results, the probability of the matter density
parameter $\Omega_m=0.300$ being consistent with the GP
reconstructed $\mathcal{D}_{\rm obs}$ function is $99.99\%$, while
the probabilities that the matter density parameter $\Omega_m=0.285$
and $\Omega_m=0.315$ being inconsistent with the current SGL
observations are $90\%$. Considering the additional assumption that
a cumulative probability of $90\%$ is considered strong evidence
against the model, we demonstrate that with 390 well-observed
galactic strong lensing systems, one can expect the matter density
parameter to be estimated with the precision of $\Delta \Omega_m
\sim 0.015$.

Now one issue which should be discussed is the comparison of our
cosmological results with those of earlier studies done using
alternative probes. First of all, based on the Planck temperature
data combined with Planck lensing, Planck Collaboration XIII (2015)
gave the best-fit parameters: $\Omega_m=0.308\pm0.012$ and
$H_0=67.8\pm0.9$ (at 68.3\% confidence level). More recently, the
best-fit values of the cosmological parameters in the flat
$\Lambda$CDM model were obtained as: ${\Omega_m}=0.255\pm0.030$ and
$H_0=70.4\pm2.5 \; \rm{kms}^{-1} \; \rm{Mpc}^{-1}$, based on the
latest observations of 41 Hubble parameter $H(z)$ at different
redshifts, which were determined from the radial BAO size method and
the differential ages of passively evolving galaxies
\citep{Aghanim}. Let us note that the matter density parameter
inferred from CMB and OHD data are highly dependent on the value of
the Hubble constant. Considering the well known strong degeneracy
between $\Omega_m$ and $H_0$. Therefore independent measurement of
$\Omega_m$ from strong lensing statistics, with the precision
comparable to \textit{Planck} observations of the CMB radiation,
could be expected and indeed is revealed here.

Working on the DGP model, the cumulative probability distributions
are plotted in Fig.~6, which also locate the $\Delta A$ values with
different matter density parameter. The probabilities associated
with differential areas are also summarized in Table~1. Similarly,
the probability of the matter density parameter $\Omega_m=0.275$
being consistent with the GP reconstructed function
$\mathcal{D}_{\rm obs}$ is $99.99\%$, while the probabilities that
the matter density parameter $\Omega_m=0.260$ and $\Omega_m=0.305$
being inconsistent with the current SGL observations are $90\%$.
Therefore, in the framework of this modified gravity theory, the
matter density parameter could be determined at the precision of of
$\Delta \Omega_m \sim 0.015-0.03$. More interestingly, benefit from
the redshift coverage of background sources in the lensing systems,
the methodology proposed in this analysis may provide improved
constraints on the DGP model, which was ruled out observationally
considering the precision cosmological observational data. Such
issue has been extensively discussed in many previous works
\citep{Wang08,Maartens10}.

However, there are several sources of systematics we do not consider
in the above analysis and which remain to be clarified for this
methodology. \textbf{As a final remark, we discuss several possible
sources of systematic errors, including sample incompleteness, the
determination of length of lens redshift bin, and the choice of lens
redshift shells, in order to verify their effect on the cosmological
constraints.} Firstly, based on the flat $\Lambda$CDM with the full
sample ($N=390$ lenses), we now estimate the systematic errors due
to statistical sample incompleteness, which could directly affect
the reconstructed function of the observable $\mathcal{D}_{\rm
obs}$. Fig.~7 shows the precision of the $\Omega_m$ parameter
assessment as a function of SGL sample size and Table 2 shows more
detailed results. One can see that, even with 50 SGL systems one can
effectively place stringent fits on the matter density in the
Universe ($\Delta \Omega_m \sim 0.05$), which furthermore
strengthens the probative power of our method to inspire new
observing programs or theoretical work in the moderate future.
\textbf{Secondly, after identifying the constraints on $\Omega_m$
obtained with the minimum acceptable $\Delta z_l=0.02$ and the
errors that it introduced, we should consider different values of
$\Delta z_l$ for examining the role $\Delta z_l$ plays in
cosmological constraints. It should be noted that the selected
length of lens redshift $\Delta z_l$ not only directly determines
the selection of simulated lensing systems, but also introduces
systematical uncertainties in estimating cosmological model
parameters. For the selection criteria of $\Delta z_l=0.01$, we
unbiasedly select a sub-sample including 150 strong lensing systems
out of the whole catalog of $N=390$ lenses. Based on this restricted
sub-sample, the constraints on $\Omega_m$ as a function of $\Delta
z_l$ are shown in Fig.~8. The results are summarized in Table 3. It
is apparent that the choice of the length of lens redshift bin,
$\Delta z_l$, which affects the derived average lens redshift for
the sample, will also play an important role in the
$\mathcal{D}_{\rm obs}$ reconstruction and thus the effectiveness of
this model-independent test. Such issue has been noted and
extensively discussed in the previous analysis, concerning the most
recent observations of early-type gravitational lenses
\citep{Yennapureddy}. Thirdly, in order to investigate the impact of
different lens shells on cosmological parameter distribution, we
also work on two additional different redshift shells at lower
redshift ($0.16<z_l<0.18$) and higher redshift ($0.73<z_l<0.75$),
which respectively generate 220 lenses in the following analysis. As
can be seen from the results illustrated in Table 4, the matter
density parameter can be estimated at the precision of $\Delta
\Omega_m\sim0.03$ and $\Delta \Omega_m\sim0.025$, respectively. We
remark here that the choice of lens redshift shells will slightly
affect the constraints on the model parameter $\Omega_m$, due to the
sample size difference between the selected sub-samples. Therefore,
our results strongly suggest that larger and more accurate sample of
the strong lensing data can become an important complementary probe
in the next decade.}

\begin{figure}
\begin{center}
\includegraphics[width=0.9\linewidth]{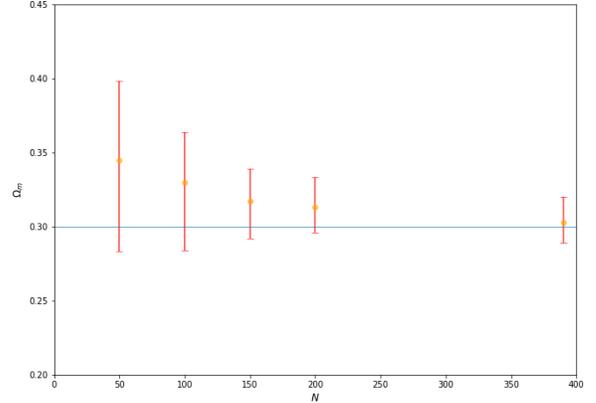}
\end{center}
\caption{Inferred $\Omega_m$ parameter shown as a function of the
number of lensing systems for $\mathcal{D}_{\rm obs}$
reconstruction.}
\end{figure}

\begin{figure}
\begin{center}
\includegraphics[width=0.9\linewidth]{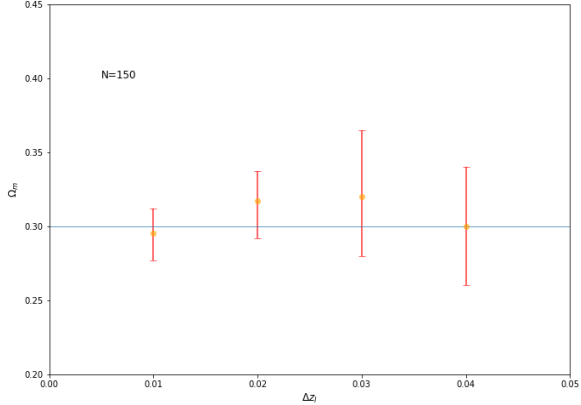}
\end{center}
\caption{\textbf{Constraints on $\Omega_m$ as a function of lens
redshift bin $\Delta z_l$. The fiducial model is shown as the dashed
line with $\Omega_m=0.30$.}}
\end{figure}

\begin{table}
\begin{center}
\caption{Summary of the cosmological constraints on $\Lambda$CDM
model with different number of lensing systems, based on 390 SGL
systems covering the redshift shell of $0.30<z_l<0.32$.}
\begin{tabular}{c|l|l}\hline\hline
Number of lensing systems     & \,\,\,\,\ Cosmological parameter  \\
\hline
   $N=50$ &\,\,\ $\Omega_m=0.345\pm0.057$   \\
  \hline
$N=100$   &\,\,\ $\Omega_m=0.330\pm0.040$   \\
  \hline
$N=150$   &\,\,\ $\Omega_m=0.317\pm0.023$   \\
  \hline
$N=200$   &\,\,\ $\Omega_m=0.313\pm0.020$   \\
  \hline
$N=390$   &\,\,\ $\Omega_m=0.300\pm0.015$   \\
\hline\hline
\end{tabular}
\end{center}
\end{table}

\begin{table}
\begin{center}
\caption{ \textbf{Summary of the cosmological constraints on
$\Lambda$CDM model with different lens redshift bin $\Delta z_l$,
based on 150 SGL systems covering the redshift shell of
$0.30<z_l<0.32$.} }
\begin{tabular}{c|l|l}\hline\hline
Length of lens redshift bin    & \,\,\,\,\ Cosmological parameter  \\
\hline
 $\Delta z_l=0.01$   &\,\,\ $\Omega_m=0.295\pm0.018$   \\
  \hline
$\Delta z_l=0.02$   &  \,\,\ $\Omega_m=0.317\pm0.023$   \\
  \hline
$\Delta z_l=0.03$   &  \,\,\ $\Omega_m=0.320\pm0.040$   \\
  \hline
$\Delta z_l=0.04$   & \,\,\ $\Omega_m=0.300\pm0.045$   \\

\hline\hline
\end{tabular}
\end{center}
\end{table}

\begin{table}
\begin{center}
\caption{ \textbf{Summary of the cosmological constraints on
$\Lambda$CDM model, based on two additional lens redshift shells.}}
\begin{tabular}{c|l|l}\hline\hline
Lens redshift shell     & Cosmological parameter   &  Probability  \\
\hline
~~~~~~~~~~~      &\,\,\,\,\,\,\,\  $\Omega_m=0.270$ &\,\,\,\,\,\,\,\  $10.00\%$   \\
$0.16<z_l<0.18$  & \,\,\,\,\,\,\,\  $\Omega_m=0.305$&    $\,\,\,\,\,\,\,\  99.99\%$ \\
~~~~~~~~~~~     & \,\,\,\,\,\,\,\  $\Omega_m=0.335$ &   $\,\,\,\,\,\,\,\  10.00\%$  \\

  \hline

~~~~~~~~~~~   & \,\,\,\,\,\,\,\  $\Omega_m=0.280$ & \,\,\,\,\,\,\,\  $10.00\%$    \\

 $0.73<z_l<0.75$   & \,\,\,\,\,\,\,\  $\Omega_m=0.300 $  &\,\,\,\,\,\,\,\  $99.99\%$    \\
~~~~~~~~~~~   &  \,\,\,\,\,\,\,\   $ \Omega_m=0.325 $   &\,\,\,\,\,\,\,\  $10.00\%$  \\

\hline\hline
\end{tabular}
\end{center}
\end{table}

\section{Conclusions}

In this paper, based on future measurements of 390 strong lensing
systems from the forthcoming Large Synoptic Survey Telescope (LSST)
survey, combined with the recently developed method based on
model-independent reconstruction approach, Gaussian Processes (GP),
we have successfully reconstructed the distance ratio
$\mathcal{D}_{\rm obs}$ reaching the source redshift $z_s\sim 4.0$.
Moreover, benefit from the Area Minimization Statistic, our results
show that independent measurement of the matter density parameter
could be expected from such strong lensing statistics at high
redshifts. Therefore, one may say that the approach initiated in
\citet{Yennapureddy} can be further developed. Here we summarize our main
conclusions in more detail:

\begin{itemize}

\item Compared with the previous statistic focusing on individual data points,
GP provides the $1\sigma$ confidence regions for the reconstructed
$D_{obs}$ function more in line with the whole sample, which greatly
restricts the possibility of cosmological models inadequately
consistent with the observational data due to otherwise large
measurement errors. However, considering the fact that our
reconstructed distance ratio $\mathcal{D}_{\rm obs}$ is an
continuous function, we apply a new statistic, the ``Area
Minimization Statistic" to constrain cosmological parameters, which
provides an effective way to make a comparison between different
models, compared with the discrete sampling statistics such as the
$\chi^2$ statistic.

\item Considering the additional assumption that a cumulative probability of 90\% is considered strong evidence
against the model, we demonstrate that with 390 well-observed
galactic strong lensing systems, one can expect the matter density
parameter to be estimated with the precision of $\Delta \Omega_m
\sim 0.015$. Such constraint is comparable to that derived from the
recent Planck 2015 results.

\item In the framework of the modified gravity theory (DGP), 390
detectable galactic lenses from future LSST survey would lead to
stringent fits of $\Delta\Omega_m\sim0.030$. More importantly,
benefit from the redshift coverage of the lensing systems, the
methodology proposed in this analysis may provide improved
constraints on the DGP model, which was ruled out observationally
considering the precision cosmological observational data. Finally,
the advantage of our method lies in the benefit of being independent
of the Hubble constant. Therefore independent measurement of
$\Omega_m$ from strong lensing statistics could be expected and
indeed is revealed here.

\item \textbf{We discuss several possible sources of systematic errors,
including sample incompleteness, the determination of length of lens
redshift bin, and the choice of lens redshift shells, in order to
verify their effect on the cosmological constraints. More
specifically, our findings indicate that the choice of the length of
lens redshift bin, $\Delta z_l$, which affects the derived average
lens redshift for the sample, plays an important role in the
$\mathcal{D}_{\rm obs}$ reconstruction and thus the effectiveness of
this model-independent test. Meanwhile, due to the sample size
difference between different selected sub-samples, the choice of
lens redshift shells will slightly affect the constraints on the
cosmological parameters.}

\item Our analysis could be extended to quantify the ability of future
measurements of strong lensing systems from the Dark Energy Survey
(DES) \citep{Frieman04}, the Joint Dark Energy Mission (JDEM)
\citep{Tyson}, and the Square Kilometer Array (SKA) \citep{McKean},
which encourages us to probe cosmological parameters at much higher
accuracy. Moreover, we also pin hope on future observational data
such as galactic-scale strong gravitational lensing systems with
Type Ia supernovae acting as background sources \citep{Cao18},
strongly lensed repeating fast radio bursts \citep{Li18}, and
strongly lensed gravitational waves (GWs) from compact binary
coalescence and their electromagnetic (EM) counterparts systems
\citep{Liao17,Cao19b,Qi19b,Qi19c}. With more detectable
galactic-scale lenses from the forthcoming surveys, the scheme
proposed in this paper can eventually be used to carry out stringent
tests on various cosmological models.

\end{itemize}

\section*{Acknowledgments}

This work was supported by National Key R\&D Program of China No.
2017YFA0402600; the National Natural Science Foundation of China
under Grants Nos. 11690023 and 11633001; Beijing Talents Fund of
Organization Department of Beijing Municipal Committee of the CPC;
the Fundamental Research Funds for the Central Universities and
Scientific Research Foundation of Beijing Normal University; and the
Opening Project of Key Laboratory of Computational Astrophysics,
National Astronomical Observatories, Chinese Academy of Sciences.


\end{document}